\documentclass[twocolumn,prl,superscriptaddress,showpacs,preprintnumbers,amsmath,amssymb]{revtex4}

\usepackage{graphicx}
\usepackage{dcolumn}
\usepackage{bm}
\usepackage{color}

\newcommand{\kb}{k_{\mbox{\scriptsize B}}}
\newcommand{\maxx}{{\mbox{\scriptsize max}}}

\begin{document}
\title{Glass Transition of the Monodisperse Gaussian Core Model}
\author{Atsushi Ikeda}
\affiliation{Institute of Physics, University of Tsukuba, Tennodai 1-1-1, Tsukuba 305-8571, Japan}
\author{Kunimasa Miyazaki}
\affiliation{Institute of Physics, University of Tsukuba, Tennodai 1-1-1, Tsukuba 305-8571, Japan}

\date{\today}
\begin{abstract}
We numerically investigate the dynamical properties of the one-component Gaussian core model in supercooled states. 
We find that nucleation is increasingly suppressed with increasing density. 
The system concomitantly exhibits glassy, slow dynamics characterized by 
the two-step stretched exponential relaxation of the density correlation and a drastic increase of the relaxation time. 
We also find a weaker violation of the Stokes-Einstein relation and a smaller non-Gaussian parameter than in typical model glass formers, 
implying weaker dynamic heterogeneities. 
Additionally, the agreement of the simulation data with the prediction of mode-coupling theory is exceptionally good, 
indicating that the nature of the slow dynamics of this ultra-soft particle fluid is mean-field-like. 
This fact may be understood as a consequence of the long-range nature of the interaction.
\end{abstract}
 \maketitle

The nature of the glass transition is surrounded by controversy. 
Several scenarios have been proposed to explain the drastic slowing down of dynamics of supercooled fluids near the glass transition point~\cite{Biroli2009,Tarjus2010,Gotze2009}. 
Numerical simulation of simple model fluids is an ideal tool to test these competing scenarios. 
However, the typical model fluids studied so far, such as Lennard-Jones, soft-core, and hard-sphere mixtures, have short-ranged, 
strong repulsive interactions in common, which dictate their thermodynamic, structural, 
and dynamical properties and render the results of these models qualitatively similar~\cite{Andersen2005}. 
A new class of model glass formers is desirable to diversify our pictures and perspectives on the glass transition within the limited accessible time windows of the simulations. 
Recently, ultra-soft particle fluids have attracted particular attention in soft-materials science~\cite{Likos2006b}. 
They are systems composed of spherical particles interacting with bounded and weak repulsions and are a good model for various soft materials, 
such as star-polymers and dendrimers. 
The absence of the hard-core-like repulsion makes the thermodynamic and 
dynamic behaviors of this class of systems extremely rich compared with standard molecular systems. 
Their phase diagrams exhibit exotic and counterintuitive properties, 
including a stable fluid phase at high temperatures for arbitrary densities, re-melting of solids at higher densities, and 
complex crystalline phases at low temperatures~\cite{Likos2006b}. 
The dynamics of the ultra-soft particles fluids also exhibits rich and nontrivial behaviors~\cite{Foffi2003b,Berthier2009f,Berthier2010e,Krekelberg2009c}.

In this Letter, we consider the simplest version of ultra-soft particles, {\it i.e.}, 
the Gaussian core model (GCM) fluid originally introduced by Stillinger~\cite{Stillinger}. 
The GCM interaction is given by 
\begin{eqnarray}
v(r) = \epsilon e^{-(r/\sigma)^2},
\end{eqnarray}
where $\epsilon$ and $\sigma$ characterize the energy and length scales, respectively. 
The GCM is an ideal model to study glassy dynamics because its thermodynamic phase diagram is relatively simple. 
Other ultra-soft particles, such as Hertzian spheres and star-polymers, exhibit complex crystalline phases, 
which may affect the dynamics in the supercooled state~\cite{Foffi2003b,Pamies2009}. 
We numerically study the {\it monodisperse} GCM in three dimension and show that nucleation is suppressed at very high densities and 
that the system exhibits canonical glassy behavior in the supercooled state. 
The quantitative agreement of the dynamical properties with the theoretical predictions is better than that of all previously investigated model glass formers.
\begin{figure}[b]
\begin{center}
\includegraphics[width=0.91\columnwidth]{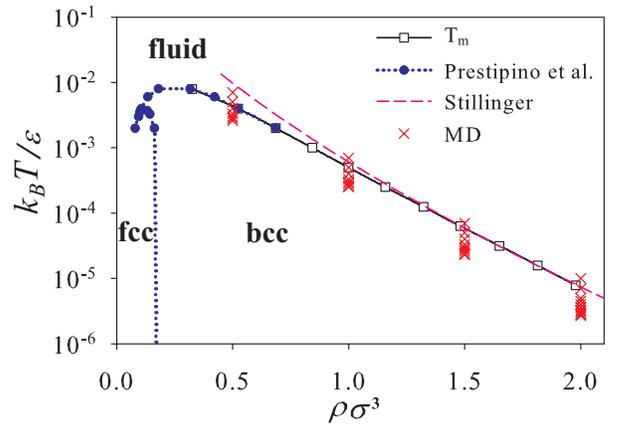}
\caption{
GCM phase diagram (empty squares). 
Results of Prestipino {\it et al.}~\cite{Prestipino2005} (filled circles) are also shown. 
The dashed line is a fit by $\log T_m \propto -\rho^{2/3}$~\cite{Stillinger}. 
The melting and freezing lines are indistinguishable at this scale. 
Crosses denote the state points where the MD simulations are performed.}
\vspace*{-0.3cm}
\label{spd}
\end{center}
\end{figure}

The thermodynamic and dynamic properties of the GCM have recently been vigorously investigated~\cite{Louis2000b,Likos2006b,Prestipino2005,Krekelberg2009c,Shall2010b}. 
Most previous studies, however, focused on the density regime $\rho\sigma^3 \lesssim 1$, 
where the phase diagram exhibits reentrant melting. 
At these densities, the monodisperse GCM easily nucleates to form crystals as it crosses the phase boundary. 
In this Letter, we investigate dynamics of the GCM near the fluid-crystal phase boundary at the unprecedentedly high densities of $\rho\sigma^3 >1$.

The thermodynamic properties of the system at high densities are carefully characterized using a Monte Carlo (MC) simulation. 
We identify the fluid-crystal phase boundary using a thermodynamic integral calculation combined with 
the particle-insertion method and the Frenkel-Ladd technique~\cite{Frenkel2001,Prestipino2005}, as shown in Figure \ref{spd}. 
Stillinger showed that the ground state of the GCM at $\rho\sigma^3 \gtrsim 0.18$ is the bcc crystal and 
argued that the melting temperature $T_m$ obeys $\log T_m \propto -\rho^{2/3}$ at high densities based on the duality relation with the hard-sphere system~\cite{Stillinger}. 
We find that $T_m$ follows this scaling at $\rho\sigma^3 \gtrsim 1$ and confirm that the crystalline structure is indeed bcc at these densities. 
The details of the thermodynamic properties will be discussed in a forthcoming paper~\cite{Ikeda_unpublished}.
\begin{figure}[t]
\begin{center}
\includegraphics[width=0.95\columnwidth]{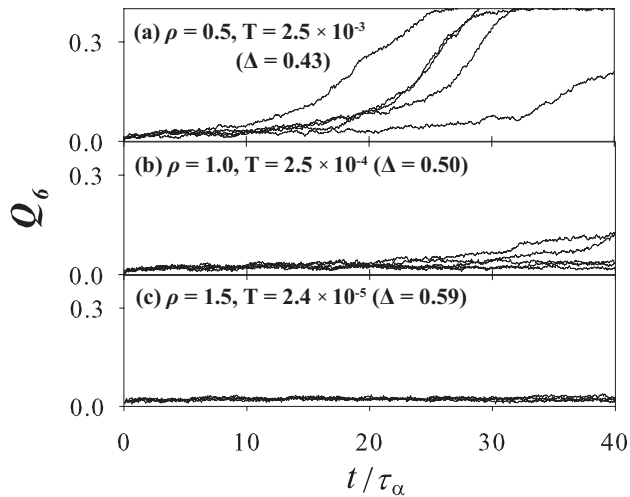}
\caption{
Time dependence of the orientational order parameter $Q_6$.
(a) $\rho = 0.5$, $T = 2.5 \times 10^{-3}$,
(b) $\rho = 1.0$, $T = 2.5 \times 10^{-4}$, and
(c) $\rho = 1.5$, $T = 2.4 \times 10^{-5}$.
$t$ is scaled by $\tau_{\alpha}$, which is evaluated from non-crystallizing samples. $\Delta \equiv (T_m-T)/T_m$ defines the distance from the melting temperature.
}
\vspace*{-0.3cm}
\label{q6}
\end{center}
\end{figure}

Dynamics of the system is investigated using a molecular dynamics (MD) simulation in the $NVT$ ensemble 
with a Nos\'{e} thermostat in the cubic cell with a periodic boundary condition. 
A time-reversible integrator, similar to the velocity-Verlet method, is used with a potential cut-off at $r=5\sigma$~\cite{Frenkel2001}. 
In the following, we take $\sigma$, $\epsilon/\kb$ and $\sigma(m/\epsilon)^{1/2}$ as the length, temperature, and time units, respectively. 
We focus on the four densities, $\rho = 0.5$, $1.0$, $1.5$, and $2.0$ 
(the melting temperatures are $T_m = 4.4 \times 10^{-3}$, $5.0 \times 10^{-4}$, $5.8 \times 10^{-5}$, and $7.2 \times 10^{-6}$, respectively) and 
perform  the MD simulations for various temperatures below $T_m$, indicated by crosses in Figure \ref{spd}. 
For each state point, five independent runs are performed to improve the statistics. The system size is fixed at $N=3456$.   
The simulations for $N=2000$ and $9826$ confirm that the finite-size effect is negligible.  
Starting from the initial configurations generated at high temperatures,  we perform the simulations for longer than $50\tau_{\alpha}$,  
where $\tau_{\alpha}$ is the alpha-relaxation time obtained from the intermediate scattering function (see below). 
The nucleation of the system into the crystalline state is monitored by the orientational order parameter $Q_6$~\cite{Steinhardt1983}. 
$Q_6$ is known to be 0.5 for the bcc crystal and zero for the disordered or fluid state~\cite{Steinhardt1983}.   
Figure \ref{q6} shows the time evolution of $Q_6$ for several trajectories at three state points. 
To compare the different states on equal footing, $t$ is scaled by $\tau_\alpha$, which is a good measure of the equilibration time. 
At $\rho=0.5$ and $T=2.5 \times 10^{-3}$, for which the distance from the phase boundary defined by $\Delta \equiv (T_m-T)/T_m$ is about 0.43, 
all five trajectories crystallize near $t \sim 20\tau_{\alpha}$ (Figure \ref{q6} (a)).  At $\rho=1.0$ and $T=2.5 \times 10^{-4}$,  
despite deeper supercooling ($\Delta= 0.5$) than that in Figure \ref{q6}(a),   
nucleation takes place at much longer times of roughly $40\tau_{\alpha}$ (Figure \ref{q6} (b)). 
For higher densities, like $\rho = 1.5$ (Figure \ref{q6} (c)), all sampled trajectories remain in the fluid state 
although the system is more supercooled ($\Delta\approx 0.6$). 
In addition to $Q_6$, we also monitor the potential energy of the system, which discontinuously decreases as the system crystallizes.  
We have checked that its time evolution is synchronized with $Q_6$. 
From these observations, we conclude that the nucleation of the GCM is strongly suppressed at very high densities.

\begin{figure}[t]
\begin{center}
\includegraphics[width=1.0\columnwidth]{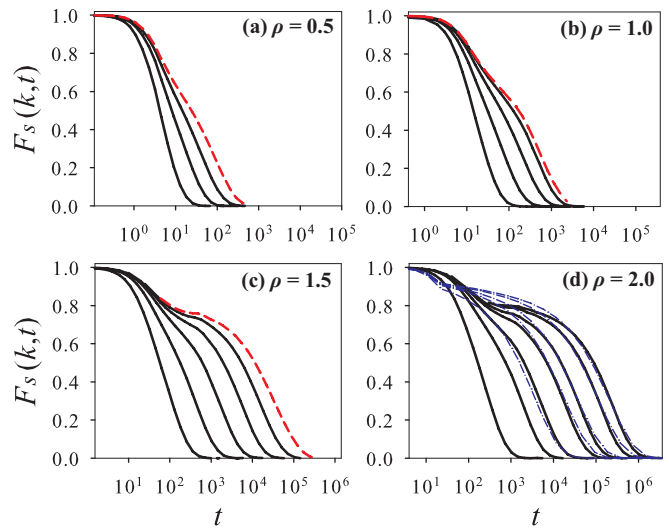}
\caption{$F_s(k,t)$ for several state points.  
(a) $\rho = 0.5$ and $T\times 10^3=7$, $4$, $3$, $2.6$, 
(b) $\rho = 1$ and $T\times 10^4=7$, $4$, $3$, $2.6$, $2.5$, 
(c) $\rho = 1.5$ and $T\times 10^5=7$, $4$, $3$, $2.6$, $2.4$, $2.3$,  and 
(d) $\rho = 2$ and $T\times 10^6=10$, $5$, $4$, $3.4$, $3.2$, $3$, $2.93$. 
The dashed lines in (a)--(c) denote the lowest-temperature data for which at least one of trajectories crystallizes.  
The dash-dotted lines in (d) are the solutions of the MCT equation.}
\vspace*{-0.6cm}
\label{fskt}
\end{center}
\end{figure}

\begin{figure}[t]
\begin{center}
\includegraphics[width=1.0\columnwidth]{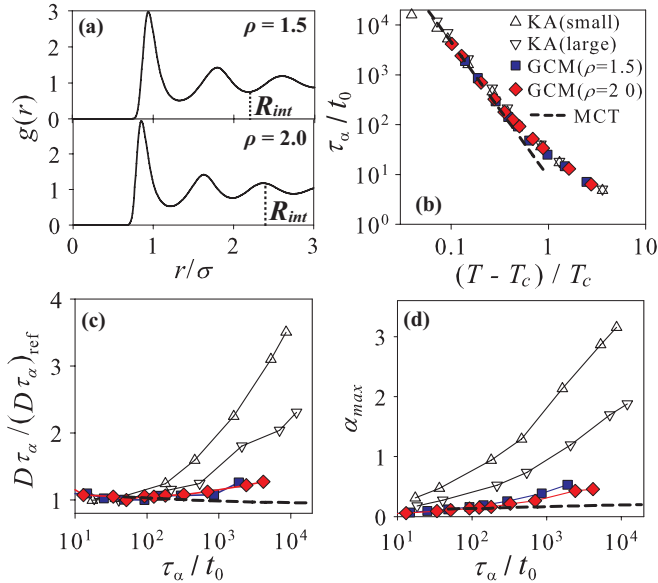}
\caption{
(a) The radial distribution function for $\rho = 1.5$ and $2.0$  at the lowest temperatures. $R_{int}$ represents the ``interaction range'' (see text).
(b) The MCT power-law fit of $\tau_\alpha$ where $T_{c}$ is a fitting parameter, 
(c) $\tau_{\alpha}$-dependence of the SE relation and (d) the peak value  of the NGP $\alpha_{\maxx}$.
Data for the small and large  particles of the KA model are also plotted.   The MCT results are shown in dashed lines.}
\vspace*{-0.6cm}
\label{alpha}
\end{center}
\end{figure}

Next, we focus on the slow dynamics of the samples that do not crystallize.  
We evaluate the self-part of the intermediate scattering function, $F_s(k,t)$, after equilibration runs of about $30 \tau_{\alpha}$. 
Figure \ref{fskt} shows the results for $F_s(k,t)$  for the wavevector $k$ near the first peak of the static structure factor $S(k)$ at various temperatures for four densities. 
Dashed lines at the lowest temperatures in Figure \ref{fskt} (a)--(c)  represent the data for the state points at which at least one of the five trajectories crystallizes. 
At the lowest density $\rho = 0.5$, crystallization always takes place before the slow dynamics sets in.  
At higher densities, however, the system clearly exhibits two-step relaxation, while the radial distribution functions $g(r)$ remain liquid-like,  
as shown in Figure \ref{alpha} (a).   
The sudden appearance of a plateau in $F_s(k,t)$ is the hallmark of the slow dynamics near the glass transition. 
The alpha-relaxation time, $\tau_\alpha$,  defined by $F_s(k,\tau_\alpha)=e^{-1}$, drastically increases as the temperature decreases. 
In particular, glassy relaxation is observed up to the lowest accessible temperature with no indication of crystallization for the densest system $\rho =2.0$.  
We also calculate the mean-square displacement  $\langle \Delta R^2(t)\rangle$ and 
observe the typical glassy behavior characterized by a plateau followed by diffusive behavior in the alpha-relaxation regime and 
the drastic decrease of the self-diffusion coefficient $D$. 

We make a more detailed characterization of the slow dynamics and compare the results with the prediction of mode-coupling theory (MCT).  
MCT has successfully described many dynamical properties of moderately supercooled fluid using $S(k)$ as the sole input~\cite{Gotze2009}.  
Though still contentious, MCT is believed to be a dynamic mean-field  theory of the glass transition~\cite{Biroli2009,Schmid2010b,Ikeda2010}. 
It predicts the relaxation behaviors of correlation functions such as  $F_s(k,t)$ semi-quantitatively and the power-law increase of $\tau_\alpha\sim |T-T_c|^{-\gamma}$,  
where $T_c$ is the temperature at which MCT predicts the spurious nonergodic transition.  
Other properties that MCT successfully predicts include the time-temperature superposition (TTS) in the alpha-relaxation regime, 
the $k$-dependence of the plateau height of the intermediate scattering function, and dynamic scaling in the plateau regime~\cite{Gotze2009}.  
On the other hand, MCT fails to capture dynamics below  $T_c$, where the activation processes over the complex energy landscape dominate.   
Another failing of MCT is that the $T_c$'s obtained by the fitting of simulation data systematically deviate from those evaluated from the theory~\cite{Kob2002}. 
Furthermore, due to the mean-field nature of the theory, 
MCT lacks an explanation of the violation of the Stokes-Einstein (SE) relation and growth of non-Gaussian parameters (NGP)~\cite{ediger2000}.  
We solve the MCT equation for the GCM using $S(k)$ obtained from simulation and compared the solution with the simulation data.  
We mainly focus on the data for $\rho=1.5$ and $\rho=2.0$, for which the plateau of the two-step relaxation of $F_s(k,t)$ is well developed. 
Our results confirm that the $F_s(k,t)$ simulation data obeys TTS  in the alpha-relaxation regime  
and can be fitted by a stretched exponential function ${e}^{-(t/\tau_{\alpha})^\beta}$ with the exponent $\beta \approx 0.7$, which agrees with the MCT results.  
The $k$-dependence of the plateau height of $F_s(k,t)$ agrees with MCT as well. 
We also find that the temperature dependence of $\tau_{\alpha}$ follows the MCT power law, 
$\tau_{\alpha} \propto |T-T_{c}^{\mbox{\scriptsize (sim)}}|^{-\gamma}$, as shown in Figure \ref{alpha} (b),  
where $\tau_\alpha$ is plotted using the short-time relaxation time $t_0$ defined by $F_s(k,t_0)=0.95$ as a time unit. 
The result for the binary Lennard-Jones system (KA model)~\cite{Kob1995} is also shown. 
We fit the data using $\gamma \approx 2.7$ (for both $\rho = 1.5$ and $2.0$), which is obtained from the MCT solution, 
and leaving $T_{c}^{\mbox{\scriptsize (sim)}}$ as the fitting parameter.  
Note that the value of $\gamma$ is comparable to that of the KA model~\cite{Kob1995}.  
These results demonstrate that the GCM shares many properties with other model glass formers.  
As shown below, however, the GCM is distinct in several respects.  First, the agreement of MCT's $F_s(k,t)$ with the simulation data is very good. 
The dash-dotted lines in Figure \ref{fskt} (d) are the MCT solution fitted using $\varepsilon = 1-T/T_c$ as the sole parameter (aside from the time unit).  
This agreement is striking, given that for other model fluids  $\varepsilon$ (and sometimes the wavevectors as well) 
needs to be adjusted at each temperature to obtain a reasonable fit~\cite{Kob2002,Voigtmann2004} (an exception is the four-dimensional system~\cite{Charbonneau2010}).  
Second, the parameters $T_{c}^{\mbox{\scriptsize (sim)}}$ used to fit $\tau_\alpha$ in Figure \ref{alpha} (b)  
are unprecedentedly close to the theoretical values $T_{c}^{\mbox{\scriptsize (theory)}}$.  
We find that $T_{c}^{\mbox{\scriptsize (sim)}} = 2.02 \times 10^{-5}$ and  $2.66 \times 10^{-6}$ for $\rho = 1.5$ and $2.0$, respectively, 
whereas their theoretical counterparts are $T_{c}^{\mbox{\scriptsize (theory)}}\!\!\!=2.66\times 10^{-5}$ ($\rho=1.5$) and $3.20\times 10^{-6}$ ($\rho=2.0$).  
The discrepancies between the simulation and theory are 32\% and 20\% for $\rho=1.5$ and 2.0, respectively. 
For other glass formers, $T_{c}^{\mbox{\scriptsize (sim)}}$ is known to differ considerably from $T_{c}^{\mbox{\scriptsize (theory)}}$. 
For the KA model, for example,  $T_{c}^{\mbox{\scriptsize (theory)}}\!\!\!\approx 0.92$ compared to  $T_{c}^{\mbox{\scriptsize (sim)}}\approx 0.44$.   
The discrepancy is more than 100\%~\cite{Kob2002}. 
The KA model at $T_{c}^{\mbox{\scriptsize (theory)}}$ is still a high-temperature fluid and $F_s(k,t)$ decays exponentially without a hint of two-step relaxation.  
On the contrary, the GCM at $T_{c}^{\mbox{\scriptsize(theory)}}$ lies deep in the region 
where the plateau of $F_s(k,t)$ is well developed (see Figure \ref{fskt}(d)). 
The third and most noticeable point is the suppression of the SE violation and the NGP.  
In Figure \ref{alpha} (c) and (d), $D\tau_\alpha$ normalized by high-temperature values  $(D\tau_\alpha)_{\mbox{\scriptsize ref}}$  
and the peak value of the NGP $\alpha_{\maxx}$ are shown as a function of $\tau_\alpha/t_0$ (instead of $T$ to compare the different systems on equal footing). 
The NGP is defined by  $\alpha(t) \equiv {3\langle \Delta R^4(t) \rangle}/{5\langle \Delta R^2(t)\rangle} - 1$.  
The variations of both $D\tau_\alpha$ and $\alpha_{\maxx}$ for the GCM are much weaker than those of the KA model~\cite{Kob1995}.  
Similar suppression of the SE violation was observed in four-dimensional systems~\cite{Charbonneau2010,Eaves2009}.  
Furthermore, $\alpha_{\maxx}$ is  smaller for $\rho=2.0$ than for $1.5$.  
Because the SE violation and the growth of the NGP are thought to be the consequences of 
the underlying dynamic heterogeneities near the glass transition point~\cite{ediger2000}, 
our results imply that the dynamic heterogeneities are weaker in the GCM
and, thus, the nature of glassy dynamics of the GCM is more mean-field-like than those of
other systems~
\footnote{One should not be confused with the so-called mean-field
approximation of the liquid state theory for the static
quantities, which predicts successfully $g(r)$fs of the GCM
at much higher temperatures than those of the present
study~\cite{Likos2006b,Louis2000b}.}.

The mean-field nature of the GCM may be attributed to 
the long-range nature of the interaction potential at high densities and extremely low temperatures, 
where many particles interact with each other.  
In Figure \ref{alpha} (a), ``interaction range'' $R_{int}$ defined by $v(r=R_{int}) = \kb T$ is indicated by dashed lines. 
$R_{int}$ reaches the second and third coordination shells, which means that many particles enter in the range of $R_{int}$ at these high densities. 
This is in stark contrast with ordinary fluid systems with strong repulsive interactions for which the interaction range is on the order  
of $\sigma$ or the distance between neighboring particles.   
A more detailed analysis supporting this hypothesis is reported elsewhere~\cite{Ikeda_unpublished}. 

An explanation of the drastic decrease of nucleation rates of the GCM at high densities is still lacking.  
It is tempting to speculate that this phenomenon is intimately  related to the mean-field character of the GCM.  
In the context of classical nucleation theory, the time scale of nucleation $\tau_n$ is proportional to that of translational diffusion $\tau_D \sim  1/D$. 
Recently, Tanaka has argued that nucleation should always intervene before the dynamic arrest takes place 
if the SE relation is violated because the decoupling of the translational motion of a single particle and structural relaxation   
leaves $\tau_n$ insensitive to temperature, while  the bulk dynamics drastically slow down~\cite{Tanaka2003b}.     
This scenario has been recently examined numerically~\cite{Saika-Voivod2009}.    
The opposite may take place for the GCM; that is, the weaker SE violation may lead to the concomitant increase of $\tau_n$ and $\tau_\alpha$, 
ultimately suppressing nucleation.

In conclusion, we demonstrated that the GCM is an unexpectedly simple and novel glass former.  
The rich dynamics of the GCM and ultra-soft particle systems in general may answer some important unanswered questions regarding the glass transition and nucleation.

\acknowledgments
 We thank Professor H. Tanaka for stimulating discussions.  A. I. is supported by the JSPS and K. M. by KAKENHI No. 2154016 and  Priority Areas ``Soft Matter Physics''.

\end{document}